\documentclass[11pt]{article}

\usepackage[lmargin=1.25in,rmargin=1.0in,tmargin=1.0in,bmargin=1.0in]{geometry}

\usepackage{epsfig}

\usepackage{sidecap}

\usepackage{color}

\usepackage[small]{titlesec}
\titlespacing{\section}{0pt}{1em}{0.25em}

\newenvironment{myindentpar}[1]
{\begin{list}{}
{\setlength{\leftmargin}{#1}}\item[]}
{\end{list}}

\begin{document}

\noindent{\Large \bf From Convection to Explosion: End-to-End Simulation of Type Ia Supernovae}

\vspace{0.25in}

\begin{myindentpar}{1in}
{\large \bf A. Nonaka$^1$,
            A. S. Almgren$^1$,
            J. B. Bell$^1$,
            H.    Ma$^2$,
            S. E. Woosley$^2$, and
            M. Zingale$^3$}

\vspace{0.1in}

$^1$Center for Computational Sciences and Engineering, Lawrence Berkeley National Laboratory, 
    Berkeley, CA 94720, USA\\
$^2$Department of Astronomy and Astrophysics, UC Santa Cruz, Santa Cruz, CA 95064, USA\\
$^3$Department of Physics and Astronomy, Stony Brook University, Stony Brook, NY 11794, USA

\vspace{0.1in}

Email: AJNonaka@lbl.gov

\vspace{0.1in}

{\bf Abstract.}  We present our end-to-end capability for computing the convective phase through
the explosion phase of Type Ia supernovae.  We compute the convective phase up to the time of
ignition using our low Mach number code, MAESTRO, and the subsequent explosion phase using
our compressible code, CASTRO.  Both codes share the same BoxLib software framework and use
finite-volume, block-structured adaptive mesh refinement (AMR) to enable high-resolution,
three-dimensional full-star simulations that scale to 100,000+ cores.  We present preliminary 
results from the first-ever simulations of convection preceding ignition using MAESTRO with 
AMR.  We also demonstrate our ability to initialize a compressible simulation of the explosion 
phase in CASTRO using data obtained directly from MAESTRO just before ignition.  Some care 
must be taken during this initialization procedure when interpreting the size and distribution 
of hot spots.

\end{myindentpar}

\vspace{0.1in}

\section{Introduction}\noindent
The most widely accepted model for a Type Ia supernova is the thermonuclear runaway and
subsequent explosion of a white dwarf that is accreting mass from a binary companion
(see \cite{hillebrandtniemeyer2000} for a review).  Two key phases of this process are the
convective phase and the explosion phase.  During the convective phase, which lasts 
hundreds of years, carbon burning near the core drives convection throughout the star along
with a gradual temperature rise.  
At some point the burning becomes vigorous enough that a hot
spot does not cool quickly enough as it buoyantly rises, and a flame front is born, i.e.,
the star ignites.  The flame propagates through the star, possibly transitioning into 
a detonation, and burns vigorously enough to cause the star to explode within a few seconds.

Previously, we have used our astrophysical code suite to compute the convective phase
and explosion phase independently.  To compute the convective phase, we use our low
Mach number code, MAESTRO \cite{MAESTRO}, and for the explosion phase, we use our
compressible hydrodynamics code, CASTRO \cite{CASTRO}.  Both codes solve the equations of
reacting flow constrained by an equation of state.
The physical processes are numerically coupled together using Strang splitting, with
a Godunov advection scheme and a pointwise ODE solver for reactions.  Additionally, 
both codes share the same BoxLib
software framework and use finite-volume, block-structured adaptive mesh refinement (AMR)
to enable high-resolution three-dimensional full-star simulations.  In \cite{SciDAC-scaling},
we described how both codes scale to 100,000+ cores using a hybrid MPI/OpenMP 
approach to parallelization.

In \cite{wdconvect}, we used MAESTRO to compute the convective phase
up to ignition at 8.7~km resolution without AMR.  We are currently performing studies
using AMR, allowing us to compute the ignition flow field at unprecedented 4.3~km resolution, 
as well as study the general flow field properties at 2.2~km.  In Section 2 we will give 
more details about the MAESTRO code, as well as give preliminary results from our newest
high-resolution simulations.

In  \cite{SciDAC-supernova} we described compressible simulations using CASTRO 
of the explosion phase at 0.5~km resolution.  We have observed, as have others, that the
characteristics of the explosion phase are highly dependent on the initial conditions, in 
particular, the size and distribution of the hot ignition points.  Unlike with previous simulations,
we are now in the unique position of being able to use simulation data from the convective phase
to initialize a simulation of the explosion, thus allowing for a true end-to-end simulation 
of a Type Ia supernova event.
In Section 3 we will provide our first observations from using MAESTRO ignition data to 
initialize a CASTRO simulation.

\section{Convection Preceding Ignition}\noindent
The convective phase preceding ignition is characterized by subsonic convection and
gradual temperature rise over hundreds of years.  We have modeled the last few hours of
convection preceding ignition using our low Mach number hydrodynamics solver, MAESTRO
\cite{MAESTRO}.  MAESTRO is based on a low Mach number equation set that 
filters acoustic waves from the system while retaining local compressibility 
effects due to reaction heating and compositional changes.  The time step constraint
in MAESTRO is based on the fluid velocity rather than the sound speed, resulting in
an average reduction in the number of time steps by a factor of $1/M$, i.e. the ratio of the sound
speed to characteristic fluid velocity.
For our simulations of convection preceding ignition, a MAESTRO time step is roughly a factor of 70 greater than 
a CASTRO time step would be.
Our scaling studies indicate that for a $576^3$ simulation with no AMR, a MAESTRO time step uses
approximately 2.5 times the computational time of a CASTRO time step, primarily due to the
linear solvers.
Thus, MAESTRO uses approximately
a factor of $\sim 70/2.5 = 28$ times less computational time than CASTRO would
to simulate the convection preceding ignition.

\begin{figure}
\centering
\includegraphics[width=1.0\textwidth]{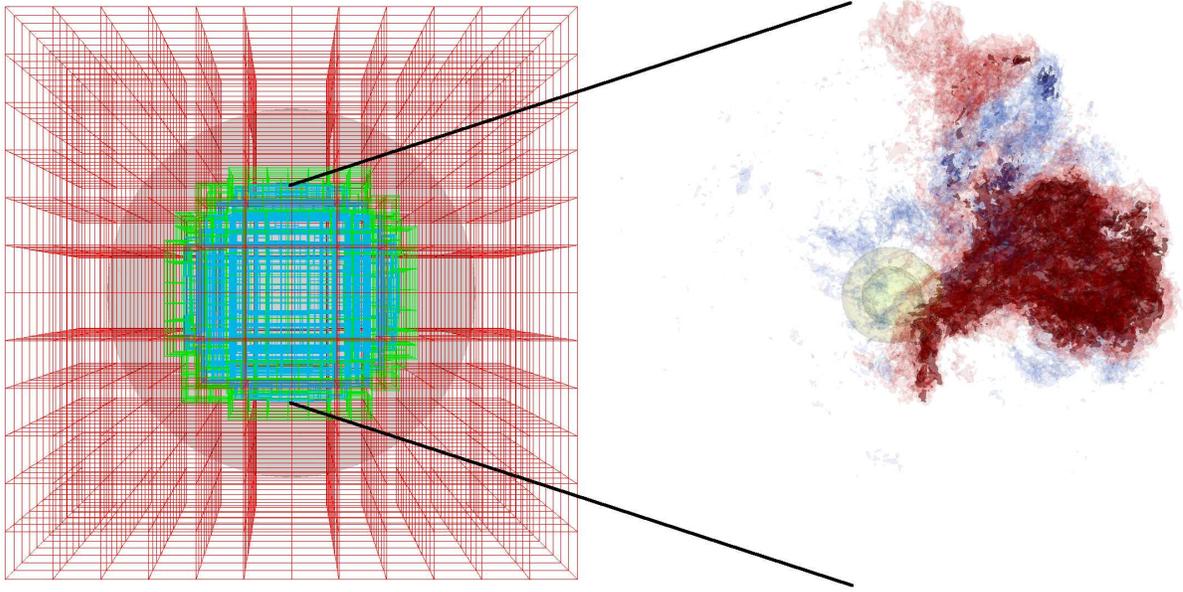}
\caption{(Left) Grid structure from a MAESTRO simulation with three total levels of AMR.
         There are $\sim 2000-4000$ grids at each level of refinement, and each grid 
         contains up to $48^3$ grid cells.
         The grids marked in blue have 2.2~km resolution cells.  (Right) Snapshot of 
         convection within the inner $\sim 1000$~km radius region around the core, which is
         marked by the blue grids.  The red/blue contours indicate regions of outward/inward 
         flow and the yellow/green contours indicate increasing burning rates.}
\label{fig:wdconvect}
\end{figure}

We define ignition in MAESTRO as the point in time when the peak temperature exceeds $8\times 10^8$~K.
In \cite{wdconvect}, we presented our results describing the convective patterns and
likely ignition radius using 8.7~km resolution simulations.  Since then, we have performed
simulations taking advantage of AMR to obtain convective patterns
with 2.2~km resolution, and ignition properties at 4.3~km resolution (computer allocations
have prevented us from running the 2.2~km case to ignition).  See Figure \ref{fig:wdconvect}
for a snapshot of the AMR grid structure and convective patterns a few minutes prior to 
ignition from the 2.2~km resolution simulation.  In this simulation, there are approximately
664 million grid cells at the finest resolution, and 1.0 billion grid cells overall.  By contrast, if
the entire domain were at the finest resolution there would be over 12 billion grid cells.

We also note the presence of a strong outward jet in the velocity field.  Our new high-resolution
simulations have led to a better understanding of the convective flow, and the behavior is more
accurately described as an outward flowing jet, rather than a dipole as we have previously
reported.  We observe that the jet changes direction over the course of the simulation.
We also noted in \cite{wdconvect} that the addition of a small amount
of rotation, included by modifying the forcing term in the momentum equation, causes the
jet to break up at lower resolution.  In the near future we will perform simulations
of rotating stars with AMR to get a clearer picture of this behavior.

In \cite{wdconvect}, we discovered that the likely ignition radius is $\sim 50-100$~km 
off-center.  It appears as if this result is robust at higher resolution, but further 
analysis is required. We also plan to analyze our data to determine the likelihood of 
multiple ignition points at different locations in the star.

\section{Explosion Phase}
\begin{SCfigure}
\centering
\includegraphics[width=0.4\textwidth]{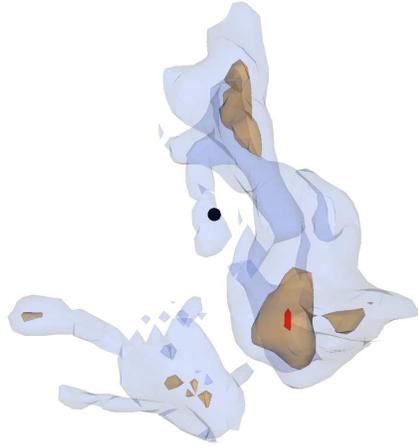}
\caption{Temperature contours at the innermost $\sim 75$~km of the white dwarf at ignition.  
         The black dot is the center of the star and has diameter 4.4~km, corresponding to 
         the cell size from the MAESTRO simulation.  The red contour encapsulates the region 
         (one grid cell in this case) where $T > 8\times 10^8$~K, the gold contours encapsulate 
         regions where $T > 7.8\times 10^8$, and the blue contours encapsulate regions where 
         $T > 7.5\times 10^8$.}
\label{fig:temperature}
\end{SCfigure}

\noindent
To model the explosion phase, we use the general compressible code, CASTRO \cite{CASTRO}.  We have
previously performed simulations of the explosion phase using artificially seeded hot
spots \cite{SciDAC-supernova}, but here we will use the data from MAESTRO directly.
We begin with the state of the star at the point of ignition from our 4.4~km MAESTRO 
simulations.  In Figure \ref{fig:temperature}, we show contours of increasing temperature near
the core at the time of ignition.  The actual ignition point in MAESTRO occurs within exactly 
one cell, which is marked by the red contour.

\begin{figure}
\centering
\includegraphics[width=0.49\textwidth]{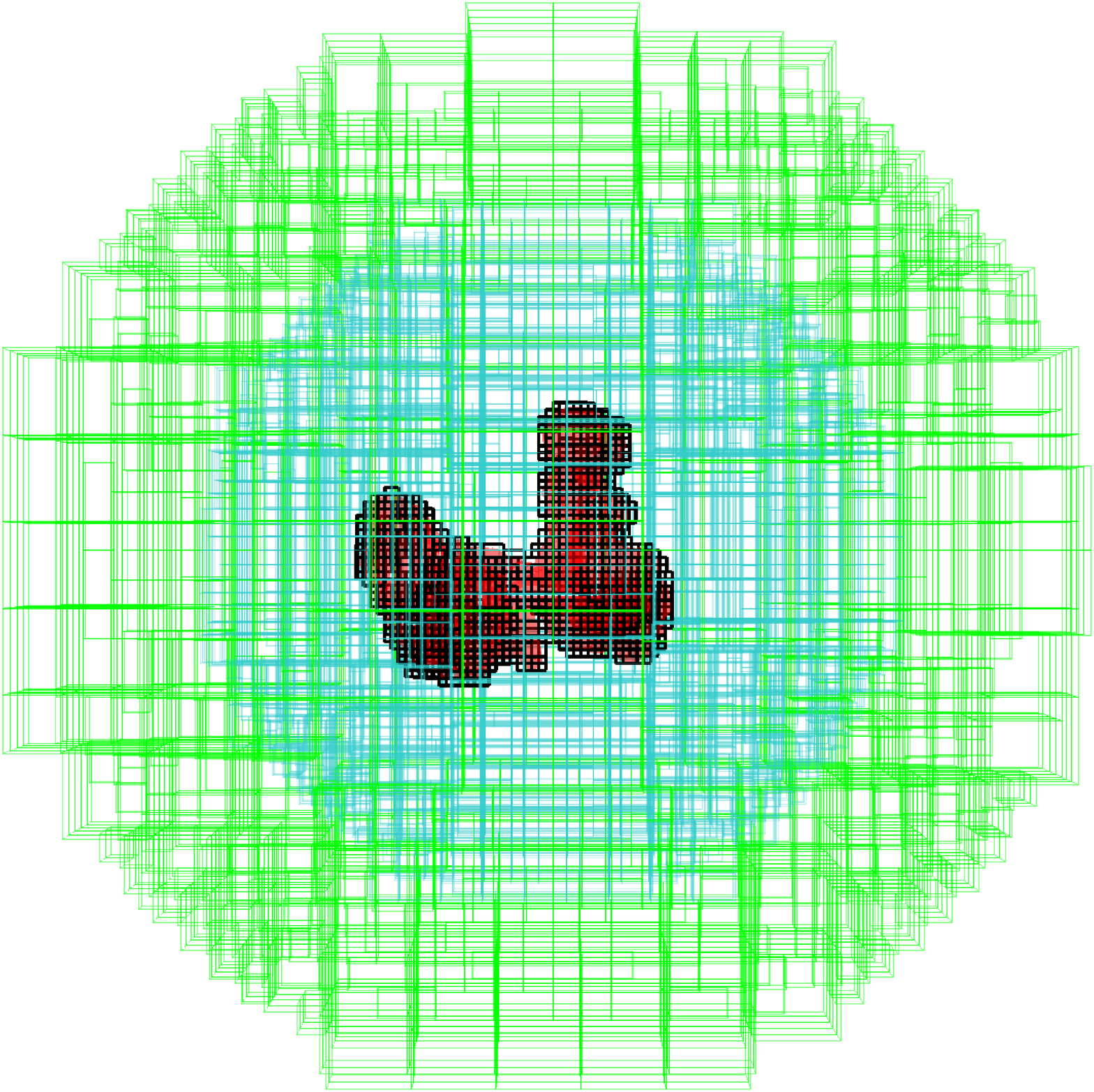}
\includegraphics[width=0.49\textwidth]{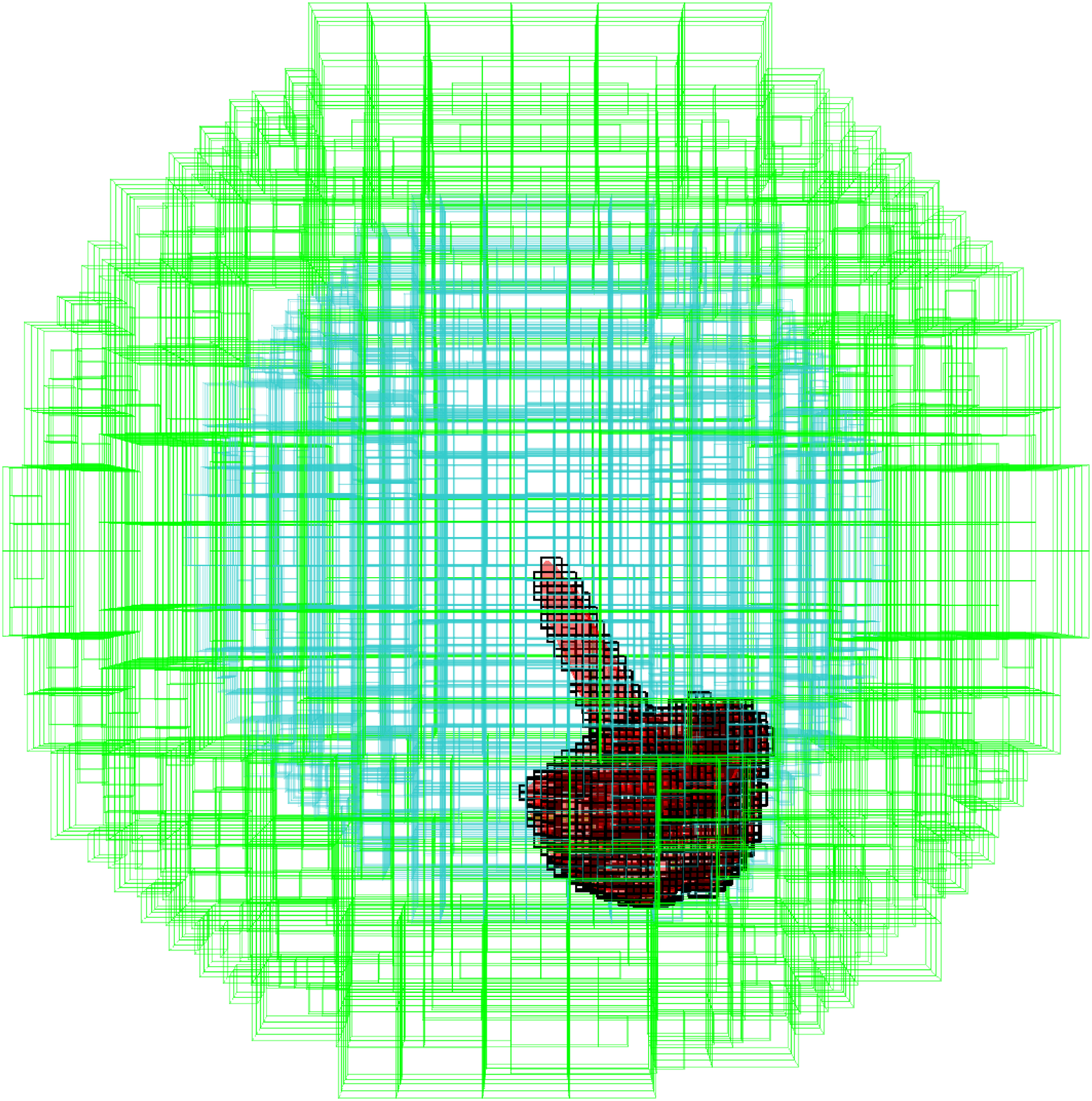}
\caption{Innermost $\sim 1500$~km of white dwarf after less than one second of evolution in 
         CASTRO using 
         two different initial conditions.  The green, blue, and black grids are at 4.3~km, 
         2.2~km, and 1.1~km resolution, respectively.  The red contour encapsulates 
         the region burned by the flame.  (Left) We initialize the flame to be in the 
         location 
         of the blue contours in Figure \ref{fig:temperature}, where $T > 7.5\times 10^8$~K.  
         (Right) We initialize the flame to be in the location of the single gold contour 
         connected to the red ignition point in Figure \ref{fig:temperature}.}
\label{fig:flame}
\end{figure}

When mapping data into CASTRO, there is some uncertainty as how to
define the ignition region, as required by the flame model in CASTRO.
Our first approach was to simply define all cells where $T > 7.5\times
10^8$~K (as indicated in blue contours in Figure
\ref{fig:temperature}) as having ignited.  The resulting flame evolved
over the first 0.5~s as shown in the left panel of Figure
\ref{fig:flame}.  Note that we have added additional resolution around
the flame front, as seen by the black grids, denoting regions where
the resolution is 1.1~km.  However, subsequent simulations in MAESTRO
in which we disabled the burning very close to the ignited cell show
that the other nearby hot spots (the gold contours in Figure
\ref{fig:temperature}), rise and cool, and therefore do not ignite in
the near future.  Based on this observation, we started another CASTRO
simulation in which we define the ignition region as the single gold
contour that is directly connected to the red ignition point.  The
subsequent flame evolution is shown in the right panel of Figure
\ref{fig:flame}.  The differences are striking, and before we continue
the evolution of the explosion phase we must be certain that we are
choosing a physically reasonable hot spot distribution.  We also note
that this flame model is preliminary and more work is needed to define
a realistic flame model for these calculations.  Nevertheless, these
simulations serve as a proof-of-concept for our capability to perform
end-to-end simulations of SNe~Ia. 


\section*{Acknowledgements}\noindent
We would like to thank Gunther Weber and Hank Childs of LBNL for their
help in using the VisIt visualization software.  The work at LBNL was 
supported by the SciDAC Program of the DOE Office of Mathematics, Information, 
and Computational Sciences under the U.S. Department of Energy under
contract No.\ DE-AC02-05CH11231.  The work at Stony Brook was
supported by a DOE/Office of Nuclear Physics grant, No.\ DE-FG02-06ER41448, to Stony Brook.
Computer time for the calculations in this paper was provided
through a DOE INCITE award at the Oak Ridge Leadership Computational
Facility (OLCF) at Oak Ridge National Laboratory, which is supported
by the Office of Science of the U.S. Department of Energy under
Contract No.\ DE-AC05-00OR22725.

\bibliography{amr_code}

\begin{thebibliography}{1}

\bibitem{SciDAC-scaling}
A.~Almgren, J.~Bell, D.~Kasen, M.~Lijewski, A.~Nonaka, P.~Nugent, C.~Rendleman,
  R.~Thomas, and M.~Zingale.
\newblock {MAESTRO, CASTRO, and SEDONA} -- {P}etascale codes for astrophysical
  applications.
\newblock {\em Proceedings of SciDAC 2010}, July 2010.

\bibitem{CASTRO}
A.~S. Almgren, V.~E. Beckner, J.~B. Bell, M.~S. Day, L.~H. Howell, C.~C.
  Joggerst, M.~J Lijewski, A.~Nonaka, M.~Singer, and M.~Zingale.
\newblock {CASTRO: A} new compressible astrophysical solver. {I. H}ydrodynamics
  and self-gravity.
\newblock {\em The Astrophysical Journal}, 215:1221--1238, 2010.

\bibitem{hillebrandtniemeyer2000}
W.~Hillebrandt and J.~C. Niemeyer.
\newblock {T}ype {I}a supernova explosion models.
\newblock {\em Annu. Rev. Astron. Astrophys}, 38:191--230, 2000.

\bibitem{SciDAC-supernova}
H.~Ma, M.~Zingale, S.~E. Woosley, A.~J. Aspden, J.~B. Bell, A.~S. Almgren,
  A.~Nonaka, and S.~Dong.
\newblock {Type Ia} supernovae: advances in large scale simulation.
\newblock {\em Proceedings of SciDAC 2010}, July 2010.

\bibitem{MAESTRO}
A.~Nonaka, A.~S. Almgren, J.~B. Bell, M.~J Lijewski, C.~M Malone, and
  M.~Zingale.
\newblock {MAESTRO: A}n adaptive low mach number hydrodynamics algorithm for
  stellar flows.
\newblock {\em The Astrophysical Journal Supplement Series}, 188:358--383,
  2010.

\bibitem{wdconvect}
M.~Zingale, A.~Nonaka, A.~S. Almgren, J.~B. Bell, and C.~M. Malone.
\newblock The convective phase preceding {Type Ia} supernovae.
\newblock {\em Submitted for publication in The Astrophysical Journal}, 2011.

\end{thebibliography}

\end{document}